\documentclass[11pt,twoside]{article}

\usepackage{asp2006}
\usepackage{epsf}
\usepackage{psfig}
\usepackage{lscape}
\usepackage{graphicx}

\bibpunct{(}{)}{;}{a}{}{,}

\begin{document}
\title{Spatial Cross Spectrum: Reducing Incoherent Convective Background of Resolved Heloseismic Instruments}   
\author{R. A. Garc\'\i a}   
\affil{Laboratoire AIM, CEA/DSM-CNRS-Universit\'e Paris Diderot; CEA, IRFU, SAp, Centre de Saclay, F-91191, Gif-sur-Yvette, France}    
\author{S. Mathur}   
\affil{Indian Institute of Astrophysics, Koramangala, Bangalore 560034, India}    
\author{I. Gonz\'alez Hern\'andez}   
\affil{National Solar Observatory, 950 North Cherry Avenue, Tucson, AZ 85719}    
\author{A. Jim\'enez}   
\affil{Instituto de Astrof\'isica de Canarias, 38205, La Laguna, Tenerife, Spain}    

\begin{abstract} 
Measurements of low-order p modes and gravity modes are perturbed by the solar convective background. Such perturbation increases below 2mHz for intensity measurements and 1mHz for velocity measurements. While the low-degree modes have large spatial scales, the convective motions have much smaller spatial distribution. In this work, we take advantage of these different scale sizes to explore the use of spatial cross spectrum between different regions of the Sun. The aim is to reduce the incoherent background noise and, therefore, increase the signal-to-noise ratio of the signals that are coherent across the full disk. To do so we use the VIRGO/LOI instrument aboard SoHO and the GONG ground-based network to study the intensity and velocity spatial cross spectra.
\end{abstract}

\section{Introduction}  
The detection of low-degree p modes and gravity modes are necessary to better constrain the structure and dynamics of the solar interior (Garc\'\i a, Mathur \& Ballot, 2008; Garc\'\i a et al. 2008b; Mathur et al. 2007, 2008).
The dominant solar background at frequencies below 2 mHz is the consequence of the surface manifestation of convection becoming an obstacle to the detection of low-degree low-order p and g modes (e.g. Garc\'\i a et al. 2001, 2004,  2007 \& 2008a). Looking at its spatial distribution we can distinguish two of these movements (e.g. Lefebvre et al. 2008): granulation motions, with horizontal scales of around 1,500 km; and supergranulation motions, with horizontal scales of around 30,000 km. We are interested in low-degree p and g modes having a few nodal lines in the solar surface. Thus, the idea is to build two time series using different combinations of pixels that cover different regions of the Sun. The evolution of the convection will be different in each time series (incoherent) while the modes will be mostly the same (high coherence between the pixels). Then, we perform the spatial cross spectrum between these two time series.
To reduce the variance of each point of the cross spectrum we take small subseries and we average all of them. We call this method the Average Spatial Cross Spectrum (AvSCS).

\section{Methods and Data Sets}
To reduce the adverse effects of noise, two independent measurements, A and B,  can be made and we can extract the common signal by using cross-spectrum techniques. The cross spectrum (CS) is defined as the complex product of the Fourier transform of one data set, A, and the complex conjugate of the Fourier transform of the other one,  B, i.e.:	
\begin{equation}
S(\nu)=\langle A(\nu) \cdot B^*(\nu) \rangle
\label{eq1}
\end{equation}

The CS enhances any coherent signals of short lifetimes improving their signal-to-noise ratio (Appourchaux et al. 2007).
The degree of coherence between the signals is measured using the coherency function, C, defined as:

\begin{equation}
C(\nu)=\frac{\langle A(\nu) \cdot B^*(\nu) \rangle}{[\langle A(\nu) \cdot A^*(\nu) \rangle \cdot \langle B(\nu) \cdot B^*(\nu) \rangle]^{1/2}}
\label{eq1}
\end{equation}

This coherency function is equal to zero for completely incoherent signals and unity for completely coherent signals. This method has already been used in helioseismology, but only by combining two contemporary datasets (Elsworth et al. 1994, Garc\'\i a et al. 1998) or by interleaving one single time series (Garc\'\i a et al. 1999). 

This study is based on the analyses of helioseismic imaged solar data. We have tested the methodology on two different types of measurements: one from LOI/VIRGO\footnote{Fr\"ohlich et al. 1995} observing the Sun using photometry (4098 days of 1 minute sampling rate from April 11, 1996 to June 30, 2007) and the other one from GONG\footnote{Harvey et al. 1996} network measuring the Doppler velocity (730 days of merged images with 1 minute cadence from January 1, 2003 to December 31, 2004).

\section{LOI Results}
Two time series were built: A, pixels={1,6,4,7} and B, pixels={2,8,3,5} (see Fig. 7 of Fr\"ohlich et al. 1995 for a full description of the pixels). In such way, we mainly preserve the same visibility to the $m$ components of the modes as in disk-integrated observations. 
 \begin{figure}[htb*]
    \centering
    \includegraphics[width=10.5cm]{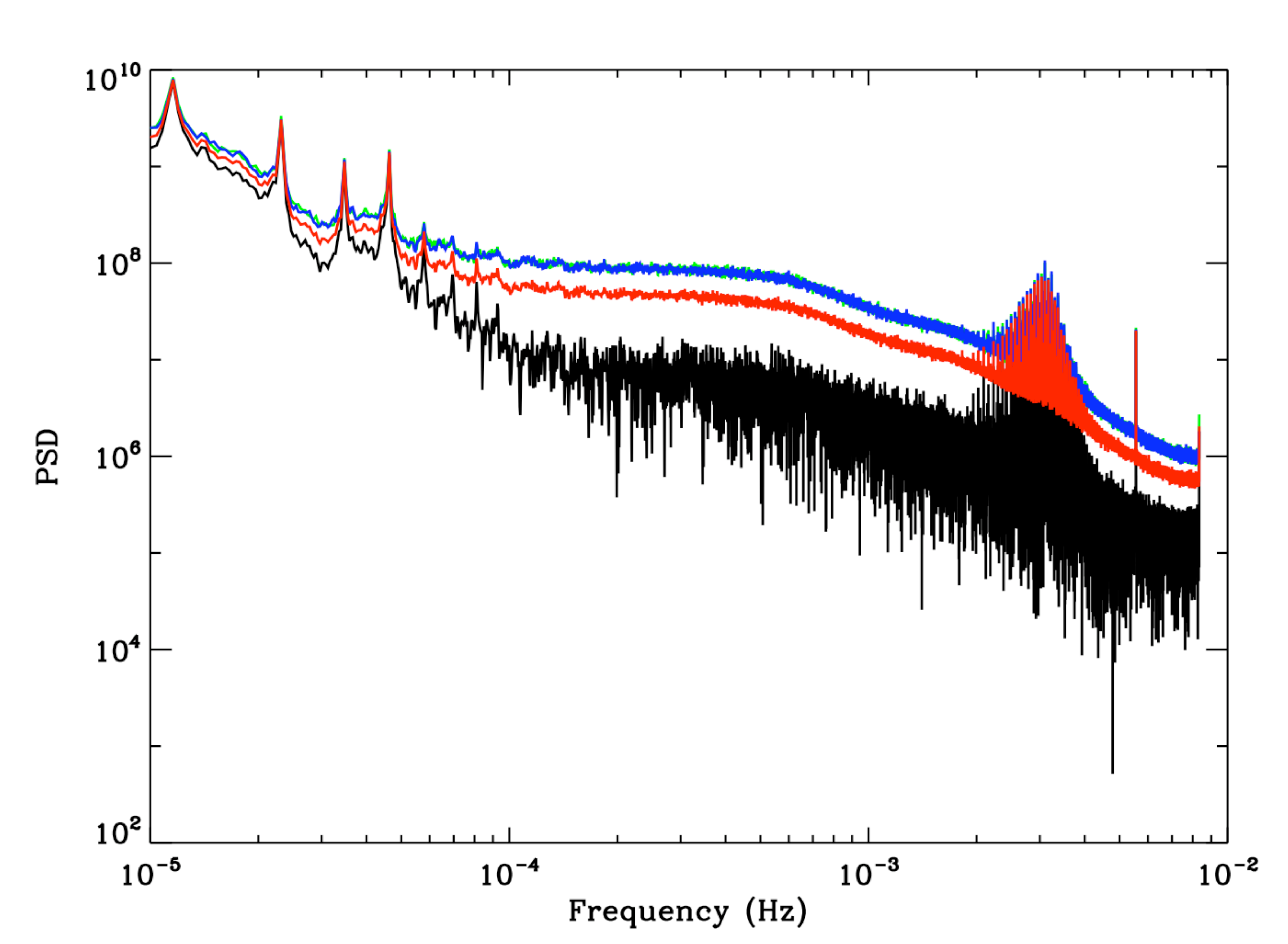}
    \caption{PSD of series A \& B (green and blue that are superimposed), PSD of the average of both channels A \& B (red) and AvSCS (black). \label{FigLOI}}
\end{figure}

We compute the AvSCS of subseries of 30 days and the result is plotted in Fig. 1. The convective background has been reduced by an order of magnitude at 1 mHz. The S/N of the p modes has been increased as well as the detection lower limit has been pushed towards lower frequencies.

\section{GONG Results}
We decomposed the disk image of GONG into an LOI-proxy image. We performed a Backwards Difference Filter (BDF) to reduce low-frequency residuals from the rotation correction as well as other merging-related effects. We corrected the power spectrum from the effect of the BDF (Garc\'\i a \& Ballot 2008).
 \begin{figure}[htb*]
    \centering
    \includegraphics[width=10.5cm]{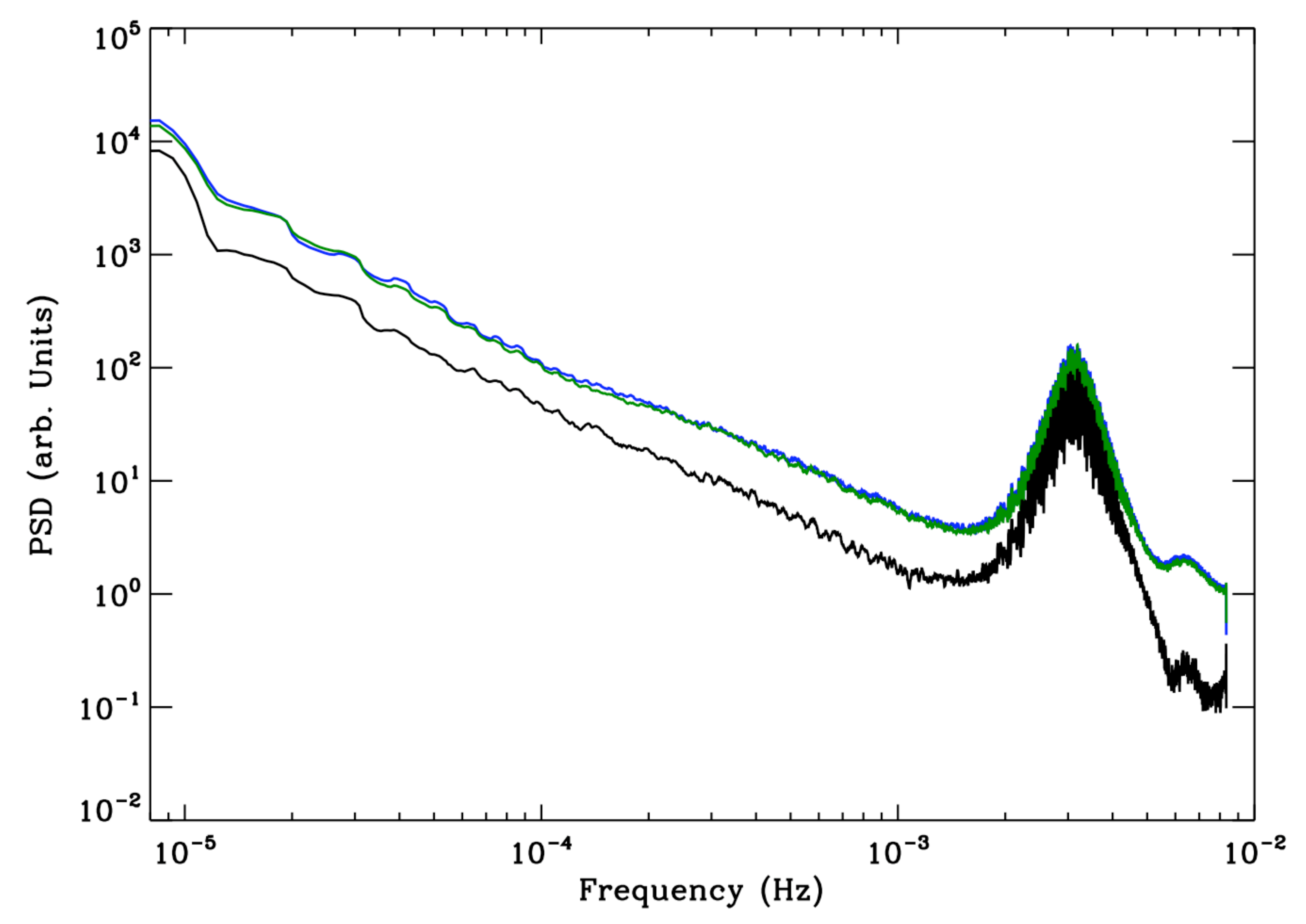}
    \caption{AvSCS (black) of 142 subseries of 15 days shifted every 5 days. The blue and green courbes are the averaged PSD of channels A and B.\label{FigGONG}}
\end{figure}
Two time series were built using the pixels of the LOI-proxy, A={9,12} and B={10,11} and the AvSCS was calculated using subseries of 15 days (shifted every 5 days). The result is plotted in Fig.2. The photon noise is reduced at high frequency as well as the convective noise at low frequency (30$\%$ coherence at 1mHz). 

\section{Conclusions}
These preliminary results show that the Averaged Spatial Cross Spectrum could be a powerful tool to reduce the incoherent noise on helioseismic measurements. However, a trade off is required between the length of the subseries, the pixels used in the independent series and the reduction of the noise obtained. In the case of LOI/VIRGO series, we have shown that the reduction of noise is quite significant (1 order of magnitude using subseries of 30 days and a total time span of $\sim$ 4100 days), while for the GONG time series an important noise reduction is also seen taking into account the reduced time series used (730 days). Longer series will be needed in the velocity measurements to better evaluate the method to detect new low-degree low order p and g modes.

\acknowledgements 
The authors want to thank the GONG Team at Tucson, for their support and very valuable discussions. This work has been partially funded by the Spanish grant PENAyA2007-62650 and the CNES/GOLF grant at the Sap-CEA/Sacaly. SOHO is an international cooperation between ESA and NASA. This work utilizes data obtained by the Global Oscillation Network Group (GONG) program, managed by the National Solar Observatory, which is operated by AURA, Inc. under a cooperative agreement with the National Science Foundation. The data were acquired by instruments operated by the BBSO, HAO, Learmonth Solar Observatory, Udaipur Solar Observatory, IAC, and Cerro Tololo Interamerican Observatory.


\end{document}